\documentclass[aps,prb,superscriptaddress, 12pt,amsmath,amssymb,longbibliography,preprint]{revtex4-2}

\usepackage{graphicx}
\usepackage{siunitx}	
\usepackage{amssymb}
\usepackage{datetime}
\usepackage{hyperref}
\usepackage[noabbrev, capitalise, nameinlink]{cleveref}
\renewcommand{\autoref}[1]{\cref{#1}}
\usepackage{xcolor}
\hypersetup{
    colorlinks,
    linkcolor={red!50!black},
    citecolor={blue!50!black},
    urlcolor={blue!80!black}
}
		
\newcommand{\TiSe}{\texorpdfstring{TiSe$_2$}{TiSe2}}
\newcommand{\CuTiSe}{\texorpdfstring{Cu$_x$TiSe$_2$}{TiSe2}}
\newcommand{\NbSe}{\texorpdfstring{NbSe$_2$}{NbSe2}}
\newcommand{\TCDW}{\ensuremath{T_{\text{CDW}}}}
\newcommand{\Tc}{\ensuremath{T_{\text{c}}}}
\newcommand{\Tcmax}{\ensuremath{T_{\text{c}}^{\text{max}}}}
\newcommand{\mstar}{\ensuremath{m^{\star}}}
\newcommand{\mband}{\ensuremath{m_{\text{band}}}}
\newcommand{\PCDW}{\ensuremath{P_{\text{CDW}}}}

\newcommand{\GPa}{\giga\pascal}
\newcommand{\kT}{\kilo\tesla}

\newcommand{\Fa}{\ensuremath{F_\alpha}}
\newcommand{\Fb}{\ensuremath{F_\beta}}
\newcommand{\Fc}{\ensuremath{F_\gamma}}
\newcommand{\Fd}{\ensuremath{F_\delta}}
\newcommand{\Fe}{\ensuremath{F_\epsilon}}

\begin{document}

\title{Lifshitz transition enabling superconducting dome around the quantum critical point in \TiSe }

\author{R. D. H. Hinlopen}
\affiliation{HH Wills Physics Laboratory, University of Bristol, Tyndall Avenue, Bristol, BS8 1TL, UK}

\author{Owen Moulding}
\affiliation{HH Wills Physics Laboratory, University of Bristol, Tyndall Avenue, Bristol, BS8 1TL, UK}
\affiliation{Institut N\'{e}el CNRS/UGA UPR2940, 25 Avenue des Martyrs, Grenoble, 38042, France}

\author{Will Broad}
\affiliation{HH Wills Physics Laboratory, University of Bristol, Tyndall Avenue, Bristol, BS8 1TL, UK}

\author{Jonathan Buhot}
\affiliation{HH Wills Physics Laboratory, University of Bristol, Tyndall Avenue, Bristol, BS8 1TL, UK}
\affiliation{High Field Magnet Laboratory (HFML-EMFL), Radboud University, Toernooiveld 7, Nijmegen, 6525 ED, The Netherlands}

\author{Femke Bangma}
\affiliation{High Field Magnet Laboratory (HFML-EMFL), Radboud University, Toernooiveld 7, Nijmegen, 6525 ED, The Netherlands}

\author{Alix McCollam}
\affiliation{High Field Magnet Laboratory (HFML-EMFL), Radboud University, Toernooiveld 7, Nijmegen, 6525 ED, The Netherlands}

\author{Jake Ayres}
\affiliation{HH Wills Physics Laboratory, University of Bristol, Tyndall Avenue, Bristol, BS8 1TL, UK}
\affiliation{High Field Magnet Laboratory (HFML-EMFL), Radboud University, Toernooiveld 7, Nijmegen, 6525 ED, The Netherlands}

\author{Charles Sayers}
\affiliation{Department of Physics, University of Bath, Bath, BA2 7AY, UK}

\author{Enrico Da Como}
\affiliation{Department of Physics, University of Bath, Bath, BA2 7AY, UK}

\author{Felix Flicker}
\affiliation{School of Physics and Astronomy, Queen's Buildings North Building, 5 The Parade, Newport Road, Cardiff, CF24 3AA, UK}

\author{Jasper van Wezel}
\affiliation{Institute for Theoretical Physics, University of Amsterdam, Science Park 904, Amsterdam, 1098 XH , The Netherlands}

\author{Sven Friedemann}
\affiliation{HH Wills Physics Laboratory, University of Bristol, Tyndall Avenue, Bristol, BS8 1TL, UK}
\email{Sven.Friedemann@bristol.ac.uk}

\date{\today}

\begin{abstract}
\textbf{
Superconductivity often emerges as a dome around a quantum critical point (QCP) where long-range order is suppressed to zero temperature. So far, this has been mostly studied in magnetically ordered materials. By contrast, the interplay between charge order and superconductivity at a QCP is not fully understood. Here, we present resistance measurements proving that a dome of superconductivity surrounds the charge-density-wave (CDW) QCP in pristine samples of $1T$-\TiSe\ tuned with hydrostatic pressure. Furthermore, we use quantum oscillation measurements to show that the superconductivity sets in at a Lifshitz transition in the electronic band structure. We use density functional theory to identify the Fermi pockets enabling superconductivity: large electron and hole pockets connected by the CDW wave vector $\vec{Q}$ which emerge upon partial suppression of the zero-pressure CDW gap. Hence, we conclude that superconductivity is of interband type enabled by the presence of hole and electron bands connected by the CDW $\vec{Q}$ vector. Earlier calculations show that interband interactions are repulsive, which suggests that unconventional $s_{\pm}$ superconductivity is realised in \TiSe\ -- similar to the iron pnictides. These results highlight the importance of Lifshitz transitions in realising unconventional superconductivity and help understand its interaction with CDW order in numerous materials.
}
\end{abstract}

\maketitle

\section{Main}
Density waves (DWs) underlie the magnetic and charge order in many materials, with well-known examples including CePd$_2$Si$_2$, BaFe$_2$As$_2$, and \TiSe\ \cite{Mathur1998,Shibauchi2014,Ishioka2010}. DWs may be formed by a periodic spatial variation of charge (charge-density wave, CDW) or spin (spin-density wave, SDW) at a characteristic wave vector $\vec{Q}$ \cite{Gruener1994}. The new periodicity breaks the translational symmetry of the underlying atomic lattice and causes a reconstruction of the electronic Fermi surface. Continuous DW phase transitions are accompanied by diverging fluctuations at $\vec{Q}$. In the vicinity of a QCP (where the transition is suppressed to zero temperature), the critical fluctuations associated with the suppressed DW order can give rise to non-Fermi liquid behaviour, and are strong contenders for mediating superconductivity in many unconventional superconductors including iron pnictides, heavy fermion materials, and possibly cuprate high-temperature superconductors \cite{Si2016, Monthoux2007, Taillefer2010}. For instance, in iron-pnictides SDW fluctuations promote unconventional $s_{\pm}$ superconductivity by coupling electron and hole pockets connected by $\vec{Q}$ \cite{Si2016}. 

A dome of superconductivity around a QCP is taken as one of the hallmarks of unconventional superconductivity. Prominent SDW systems demonstrate such a dome, including the heavy fermion compound CePd$_2$Si$_2$ \cite{Mathur1998}, iron-pnictide BaFe$_2$As$_2$ \cite{Shibauchi2014}, and the organic superconductor (TMTSF)$_2$PF$_6$ \cite{Vuleti2002}. This is in clear contrast to simple competition between DW order and superconductivity as reported for instance in \NbSe, where the dome of superconductivity is absent \cite{Moulding2020}. Complex interaction between CDW order and superconductivity has been observed in underdoped \cite{Putzke2018} and overdoped \cite{Tam2022} cuprates and nickelates \cite{Tam2022a} as well as in frustrated and topologically non-trivial AV$_3$Sb$_5$ (A=K, Rb, Cs) \cite{Ortiz2020}. We report that \TiSe\ stands apart from all these by being the cleanest example in which a superconducting dome surrounds the CDW QCP.

\begin{figure*}
\includegraphics[width=\textwidth]{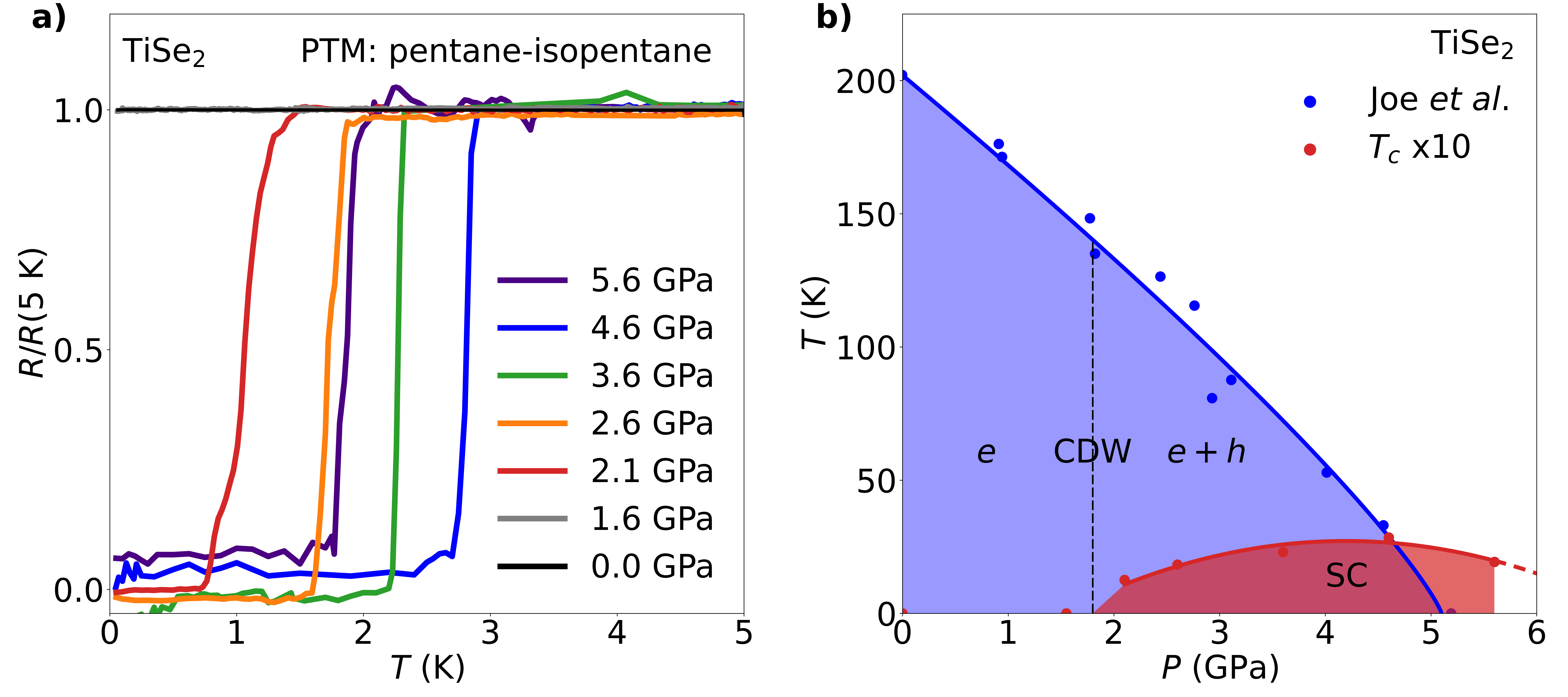}
\caption{\textbf{Superconductivity around the CDW QCP in \TiSe.} (a) Resistance measurements using a hydrostatic pressure transmitting medium (PTM) demonstrate superconductivity over an extended pressure range in \TiSe. (b) Superconducting transition temperatures \Tc\ defined as the temperature at which the resistance is reduced to \SI{90}{\percent} of the normal state resistance. The dashed vertical line indicates the pressure at which we observe a Lifshitz transition. \TCDW\ is also shown as obtained from high-pressure X-ray diffraction measurements (blue points) using a similarly high-quality pressure medium \cite{Joe2014, Moulding2022}. Solid lines and shading are guides to the eye. } 
\label{fig:SC_dome}
\end{figure*}

In this article, we report on the interplay between CDW order, superconductivity, and Lifshitz transitions in the transition metal dichalcogenide $1T$-\TiSe. A Lifshitz transition marks a topological change in the Fermi surface \cite{Lifshitz1960}. \TiSe\ hosts a prototypical CDW associated with a doubling of the unit cell in all directions ($\vec{Q}=\left(\frac{1}{2}, \frac{1}{2},\frac{1}{2}\right)$) and a transition temperature of $\TCDW=\SI{202}{\kelvin}$. Evidence exists for an excitonic contribution to the CDW mechanism \cite{Jerome1967, Rossnagel2011, Hellmann2012,Abbamonte2017} and the CDW may be chiral \cite{Ishioka2010}. Previous studies found the CDW is suppressed and superconductivity can be induced by Cu intercalation  \cite{Morosan2006}, high pressure \cite{Kusmartseva2009, Joe2014}, or gating \cite{Li2015}. Combined with its structural simplicity and lack of magnetic order, \TiSe\ thus provides an ideal setting to investigate the emergence of superconductivity around a CDW QCP.
Under hydrostatic conditions, the CDW transition is continuously suppressed to zero temperature at a pressure of $\PCDW=\SI{5}{\GPa}$, as observed by X-ray diffraction and magnetotransport measurements \cite{Joe2014,Moulding2022}. Hence, this provides evidence for a CDW QCP at \PCDW.
Yet, the link between superconductivity and CDW order in \TiSe\ remains elusive, as the exact location of superconductivity relative to the CDW QCP remains debated \cite{Morosan2006, Kusmartseva2009,Moulding2022,Joe2014, Kogar2017a,Saqib2021,Xia2022,Lee2021a}.

We study the intrinsic behaviour of pristine samples by using hydrostatic pressure to fully suppress the CDW phase of \TiSe. We measure the low-temperature resistance to map the superconducting dome, observe quantum oscillations across the full CDW phase and beyond, and combine these results with density functional theory (DFT) and a tight-binding analysis to map the evolution of the electronic structure identifying two Lifshitz transitions. We find that the onset of superconductivity coincides with the emergence of a major hole and electron pocket at one of the Lifshitz transitions. Combined with earlier theoretical work \cite{Wezel2011},  this suggests that \TiSe\ hosts unconventional superconductivity with interband pairing at the CDW $Q$-vector, and may be the first case of superconductivity with $s_{\pm}$ character in a CDW compound.

\section{Superconductivity at the CDW QCP in {\TiSe}}

Our resistance measurements on pristine samples tuned with hydrostatic pressure show that the superconducting transition temperature \Tc\ forms a dome around the CDW QCP (see \autoref{fig:SC_dome}). While \TiSe\ remains a normal metal at ambient and low pressures down to at least 60 mK, we find sharp superconducting transitions at pressures $P>$~\SI{2.0}{GPa}.  Previously, superconductivity was observed between \SI{2.0}{\GPa} and \SI{3.5}{\GPa} with a maximum transition temperature $\Tcmax=\SI{1.8}{K}$, in a study using a solid pressure medium \cite{Kusmartseva2009}. By using the liquid pressure transmitting medium (PTM) 1:1 pentane-isopentane, we improve on the hydrostatic conditions. In hydrostatic conditions, we find that superconductivity still sets in at \SI{2}{\GPa}, but now extends to at least \SI{5.6}{\GPa}, with an enhanced maximum transition temperature $\Tcmax=\SI{2.9}{K}$ close to the CDW QCP. The presence of superconductivity beyond \PCDW\ rules out the earlier suggestion \cite{Joe2014} that superconductivity is confined to the domain walls of the CDW and requires new understanding.

The comparison with previously published results underlines the sensitivity of both superconductivity and CDW order to pressure conditions. Our study shows a \SI{50}{\percent} higher \Tcmax\ in hydrostatic conditions than previous studies using solid pressure media \cite{Kusmartseva2009, Saqib2021, Xia2022}. This sensitivity of the superconductivity to non-hydrostatic conditions hints at an unconventional mechanism with sign change of the superconducting gap. A similar sensitivity to pressure conditions has been summarised for the CDW order: under the best hydrostatic conditions, the critical pressure of the CDW order is enhanced by more than \SI{50}{\percent} and  reaches $\PCDW=\SI{5}{\GPa}$ \cite{Joe2014, Moulding2022} .

The observed dome of superconductivity around the CDW QCP in \TiSe\ raises the question to what extent the vicinity of the CDW QCP influences its normal superconducting properties? Superconductivity is conventionally mediated by phonons, whilst in many unconventional superconductors magnetic or nematic fluctuations are suggested to contribute to or even dominate the binding of Cooper pairs \cite{Monthoux2007}. In heavy-fermion and iron-pnictide superconductors, these magnetic and nematic fluctuations are believed to arise from a QCP \cite{Si2016}. In \TiSe, the fluctuations around the CDW QCP involve both the lattice and the electrons, and possibly excitons \cite{Jerome1967, Rossnagel2011, Hellmann2012,Abbamonte2017}. The presence of the QCP and its associated fluctuations is known to influence the electrons, as manifested for \TiSe\ in the normal-state resistivity $\rho=\rho_0+A T^n$, which has a dip in the exponent $n$ in the vicinity of the QCP~\cite{Moulding2022}. 
The value $n=3$ away from the QCP is consistent with interband scattering driven by phonons \cite{Wilson1938}. The reduction to $n\approx 2$ near the QCP can be interpreted as evidence for an increase of the effective electron-phonon coupling $\lambda$ \cite{Gurvitch1986}. Notice that the influence of disorder, which can also cause a dip in $n$, can be ruled out in this case as we maintain constant chemical purity with pressure and observe quantum oscillations at all pressures studied. Hence, the dip in $n$ suggests that the conduction electrons predominantly couple to phonon modes that are softened to zero energy at the CDW QCP \cite{Knowles2020}. Extending this reasoning, the maximum value \Tcmax\ of the superconducting transition temperature at the QCP may be similarly enhanced by an increased peak in $\lambda$. This is consistent with a reduced dip of $n$ in previous studies, with non-hydrostatic conditions, suggesting a smaller $\lambda$ underlying the reduced \Tcmax\ \cite{Kusmartseva2009,Saqib2021,Xia2022}. 

The effective quasiparticle mass \mstar\ is expected to be enhanced over the band mass \mband\ due to electron-phonon coupling as $\mstar=(1+\lambda)\mband$. We  observe an enhancement of \mstar\ for \Fa\ rising from $\approx\SI{20}{\percent}$ to $\approx\SI{200}{\percent}$ on approaching \PCDW. 
We also observe an enhancement of \mstar\ by $\approx\SI{80}{\percent}$ for \Fd\ and \Fe\ (cf. black and purple symbols compared to same color lines). Whilst an increase of the mass enhancement is thus only seen for \Fa, the expected change in $\lambda$ may be small enough to be non-discernible for \Fd\ and \Fe\ (compare for instance \cite{Raymond2011}).

This mechanism of enhancing \Tc\ is similar to that observed in elemental uranium and LuPt$_2$In, in which an increase of \Tc\ correlates with the softening of phonons as observed by inelastic X-ray scattering and specific heat measurements, respectively \cite{Raymond2011, Gruner2017}. Yet, \TiSe\ remains the first case of a dome of superconductivity around a CDW QCP where we can study the onset of superconductivity. This allows us to answer the question: What initiates the superconductivity in \TiSe\ at \SI{2}{\GPa}? We employ quantum oscillation measurements and electronic structure calculations to address this question.

\section{High-pressure quantum oscillation measurements}

\begin{figure*}
\includegraphics[width=\textwidth]{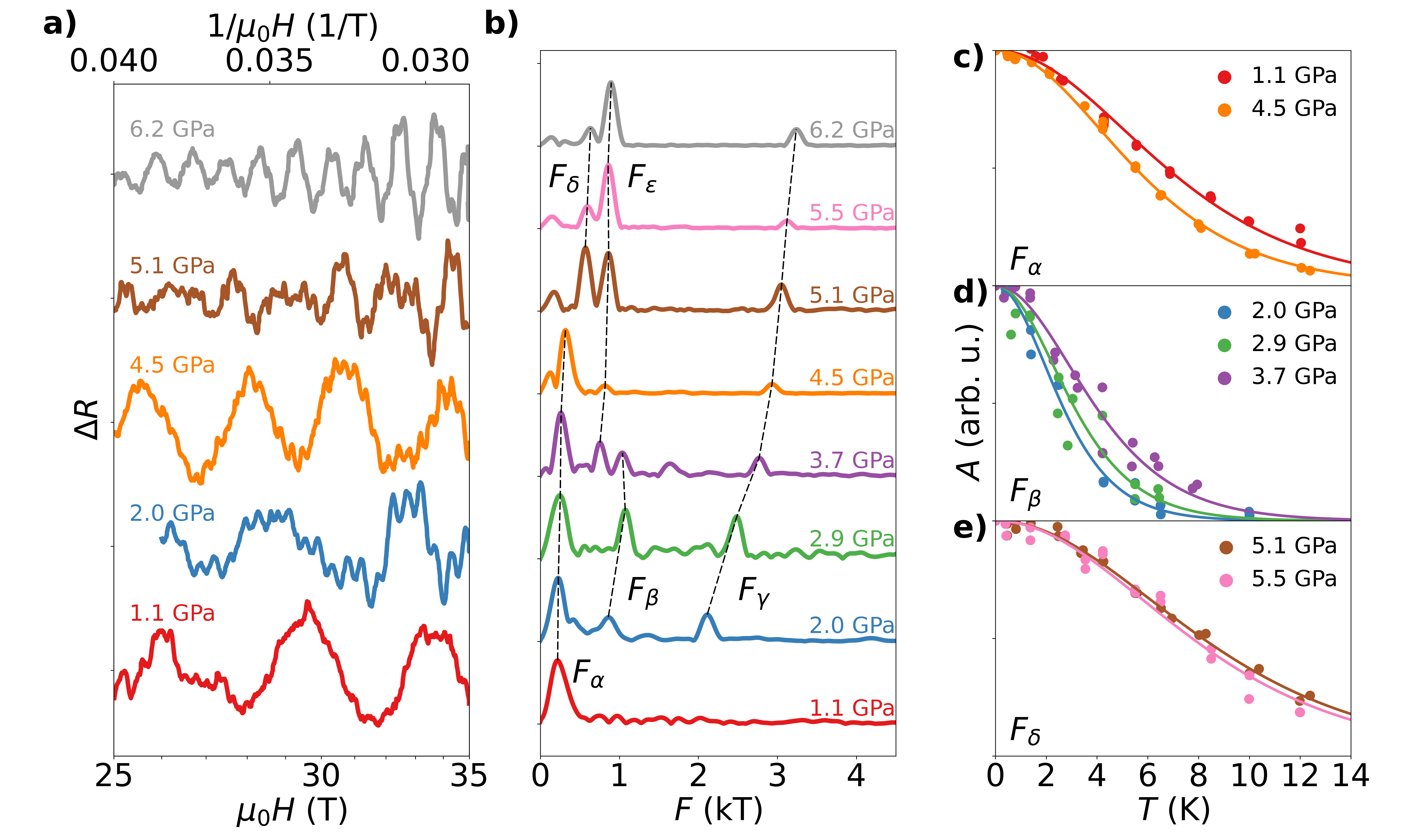}
\caption{\textbf{Quantum oscillations in \TiSe\ at high pressures.} (a) Resistance after subtraction of a polynomial background plotted versus inverse magnetic field (top axis) across a wide range of pressure. (b) Fourier transformation of the data gives the oscillation amplitude as a function of frequency. Dashed lines trace quantum oscillation frequencies across pressure. (c) The temperature dependence of the amplitudes for selected frequencies and pressures. Solid lines represent Lifshitz-Kosevich fits used to extract effective masses.}
\label{fig:QO_data}
\end{figure*}

The presence of quantum oscillations in our \TiSe\ samples presented in \autoref{fig:QO_data} demonstrate that we retain the high purity of our samples under hydrostatic pressure tuning \cite{Knowles2020}. Analysing the quantum oscillation measurements affords us the highest resolution and most reliable method to obtain information about the Fermi surface and electronic structure. Previous quantum oscillation measurements have been decisive for studies of  electronic structure and Fermi-surface reconstruction as well as the quasiparticle renormalisation at quantum phase transitions in a range of materials \cite{Shishido2005,Ramshaw2015,Semeniuk2022a}.

\begin{figure*}
\includegraphics[width=0.8\textwidth]{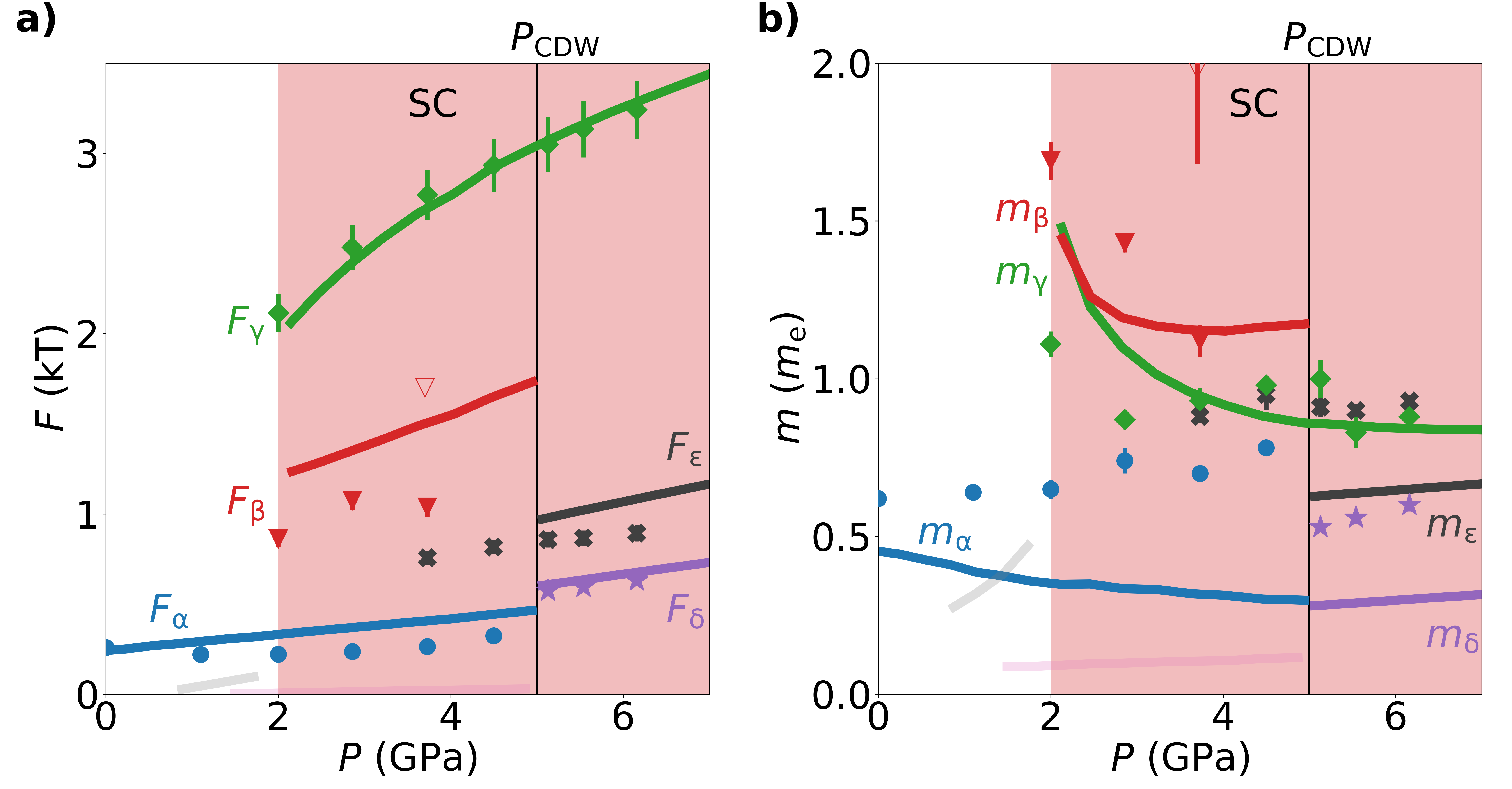}
\caption{\textbf{Comparison between modelled and measured quantum oscillations}. a) Quantum oscillation frequencies from experiment (symbols) and band-structure calculations (lines). 
Labels refer to frequencies identified in \autoref{fig:QO_data}, with subscripts corresponding to orbits in \autoref{fig:theory}. Faint lines correspond to small frequencies which are not observed experimentally. Neck orbits of the Se band at $k_z=\pi/2$ are omitted as they are not observed experimentally. b) Corresponding effective masses as a function of pressure. Grey background indicates range of superconductivity observed. Open triangle may be the result of magnetic breakdown, as discussed in Supplementary Information D.}
\label{fig:comparison}
\end{figure*}

At ambient pressure and low temperature, \TiSe\ is fully described by a single electron pocket, which manifests as a single quantum oscillation frequency $\Fa=\SI{0.26}{\kT}$ \cite{Knowles2020}. This pocket is the result of Fermi surface reconstruction inside the CDW phase and was previously observed using ARPES \cite{Watson2019}. In particular, the agreement of quantum oscillations with heat capacity data establishes the fact that \Fa\ corresponds to the \textit{only} Fermi surface pocket present at ambient pressure \cite{Craven1978}. We observe no other orbits up to \SI{2}{\GPa}. Within this pressure range, \Fa\ decreases slightly and the mass evolves smoothly (see \autoref{fig:QO_data} and \autoref{fig:comparison}).

The sudden emergence of new frequencies at higher pressures provides evidence for two Fermi surface reconstructions taking place at \SI{2}{\GPa} and \SI{5}{\GPa} (see \autoref{fig:comparison}). 
At \SI{2}{\GPa} we identify two new frequencies $\Fb\approx\SI{1}{\kT}$ and $\Fc\approx\SI{2}{\kT}$ (\autoref{fig:QO_data}). These are one order of magnitude larger than \Fa\ and prove the emergence of at least one large new Fermi surface (notice that heat capacity measurements rule out the presence of such a large pocket at ambient pressure). The emergence of such a large Fermi surface is surprising, as it happens well before the critical pressure at which CDW order disappears.

A second Fermi surface reconstruction is observed at $\PCDW=\SI{5}{\GPa}$. Here, \Fa\ ceases to exist and a new frequency $\Fd\approx\SI{0.55}{\kT}$ emerges. \Fa\ and \Fd\ are clearly distinguished by the difference in mass, establishing these as orbits on different Fermi surface pockets. By contrast,  \Fc\ evolves smoothly through $\PCDW$ and continues to increase up to the highest pressure of our study (\SI{6.2}{\GPa}). 
The electronic structure calculations described below model the complex evolution of the Fermi surfaces detected by quantum oscillations, including the two Fermi surface reconstructions. 

\begin{figure*}
\includegraphics[width=\textwidth]{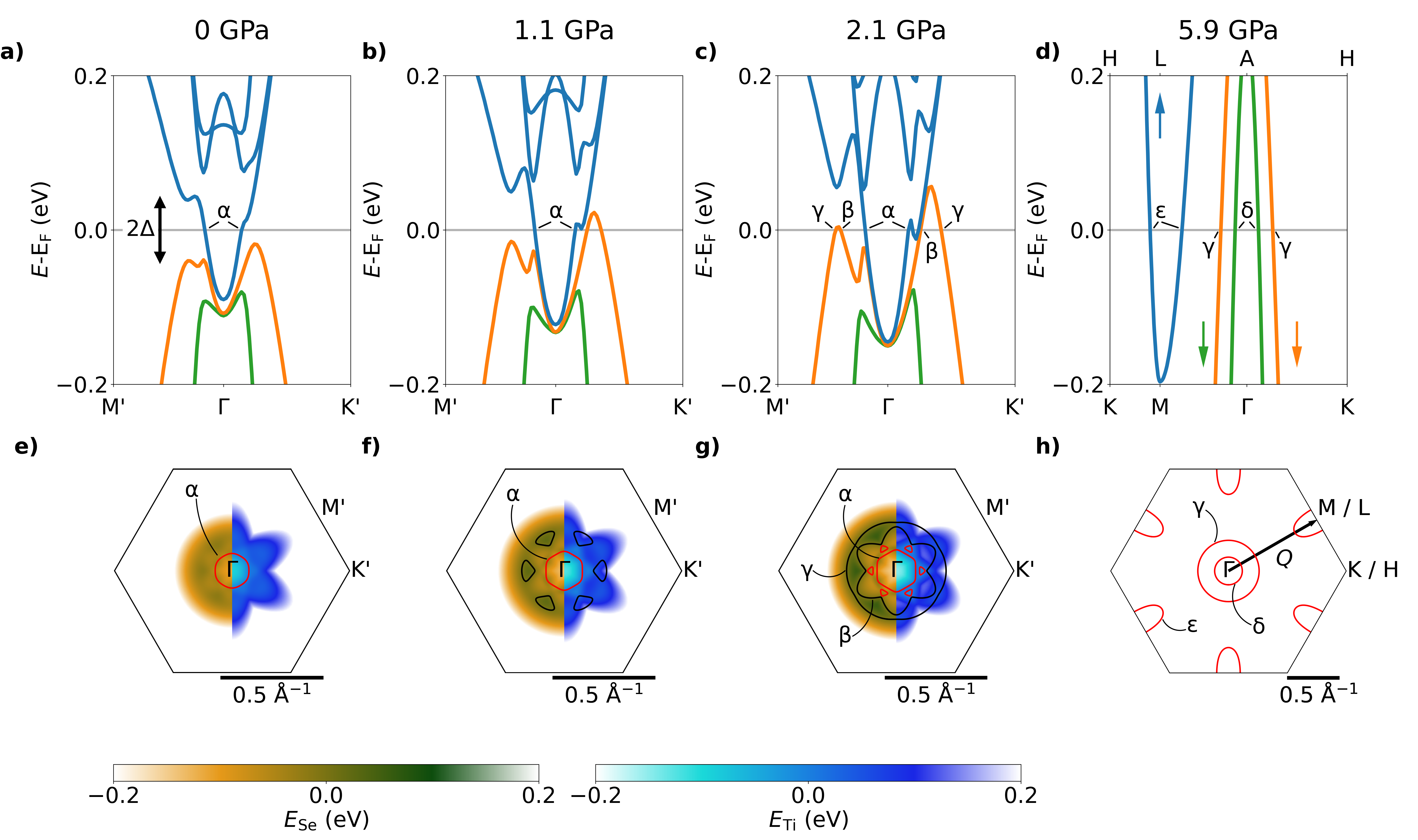}
\caption{\textbf{Pressure evolution of the \TiSe\ Fermi surface}. a-d) Simulated band structure inside (a-c) and outside (d) the CDW phase. Blue indicates tight binding fits to the Ti-3$d$ electron band back-folded from the L point, green and orange the Se-4$p$ hole bands. e-h) Corresponding Fermi surface contours in the tight-binding model. The color maps in the left half of each Brillouin zone indicate the energies of the Se tight-binding band with red lines marking the corresponding Fermi surface. The right half of each Brillouin zone with black lines correspond to the Ti bands. Orbits observed in experiment are labeled with Greek letters. h) Normal state Fermi surface outside the CDW regime, with one of the three CDW wavevectors $\vec{Q}=\left(\frac{1}{2}, \frac{1}{2}, \frac{1}{2}\right)$ indicated by the black arrow. We label points on the high-pressure Brillouin zone with $\Gamma,M, L,K,H$ where $L$ is mapped onto $M$ and $H$ onto $K$ in our two-dimensional model.  Points in the reconstructed  Brillouin zone inside the CDW phase are labelled with $\Gamma,M',K'$.} 
\label{fig:theory}
\end{figure*}

\section{Electronic structure calculations}

We start our analysis of the observed quantum-oscillation frequencies by employing DFT calculations in WIEN2k \cite{Blaha2019} including spin-orbit coupling and based on the well-established crystal structure of \TiSe\ without the CDW (space group $P\bar{3}m1$). The pressure dependence of lattice parameters is calculated with DFT and matches the experimental results (see Methods and Supplementary Information A). The DFT band structure is then used to unambiguously assign the quantum oscillation frequencies above \PCDW\ (\autoref{fig:comparison}), and to calibrate band shift corrections of the electronic structure in the absence of CDW order as detailed in Supplementary Information A. 

Outside the CDW phase ($P\geq\PCDW$), DFT calculations provide an excellent match to the experimental frequencies. The three experimental frequencies \Fc, \Fd, and \Fe\ are reproduced in both magnitude and pressure dependence. Based on this, we can clearly identify \Fc\ and \Fd\ to correspond to the outer and inner hole pockets at $\Gamma$, while \Fe\ corresponds to the electron pocket at $L$ (cf. \autoref{fig:theory}). The calibration of band shifts in the DFT calculations (see Methods and Supplementary Information A) to these observed frequencies provides us with a reliable basis to model the effect of the CDW using a tight-binding approach.

\subsection{Pressure dependence}

Inside the CDW phase ($P\leq\PCDW$), we model the Fermi surface by employing a two-dimensional tight-binding model with band parameters extracted from our calibrated DFT calculations. We include a simple $k$-independent orbitally non-selective CDW gap $\Delta$ between the Se and Ti bands (see Methods and Supplementary Information A). The gap size $\Delta(P)$ is taken proportional to $\TCDW(P)$ at all pressures, with an estimate of $\Delta(0)=\SI{40}{\milli\electronvolt}$ at ambient pressure based on \TiSe\ being a moderately strong coupling CDW with a transition temperature of $\TCDW=\SI{202}{\kelvin}$ (see Methods). Our model quantitatively accounts for the observed quantum oscillation frequencies over the full  pressure range studied, as shown in \autoref{fig:comparison}. 

The tight-binding model naturally reproduces the Fermi surface reconstruction at $\PCDW=\SI{5}{GPa}$ identified by \Fd\ outside the CDW and \Fa inside the CDW phase, as shown in \autoref{fig:comparison}. The tight binding model also predicts the disappearance of \Fe\ and emergence of \Fb\ when entering the CDW phase from high pressures. In experiment, we observe \Fe\ to continue below \PCDW\ and \Fb\ is only detected below $\SI{4}{GPa}$. The fact that \Fb\ is not observed in experiment right up to \PCDW\ may be due to the low intensity of this frequency as a result of either geometrical factors or reduced quasiparticle weight. The continued observation of \Fe\ into the CDW phase as well as the additional frequency observed at \SI{3.7}{\GPa} are possibly results of magnetic breakdown as discussed in Supplementary Information D. The new orbits \Fa\ and \Fb\ inside the CDW phase are identified as the inner and outer closed orbits around $\Gamma$ arising from the hybridisation between the three elliptical electron bands originally at $L$, and the smaller hole band originally at $\Gamma$. In our model, the $\alpha$ pocket persists down to ambient pressure, where it encompasses the entire Fermi surface in excellent agreement with experiments \cite{Knowles2020}. 
While the calculated \Fb\ is seen in \autoref{fig:comparison} to match the experimental pressure trend, the \SI{20}{\percent} difference in value between model and experimental data is a result of the limited expansion of the employed tight binding model (see Supplementary Information C). Similarly, we assign the difference between observed and calculated frequencies for \Fa\ to the limitations of our model. Indeed, quantitative prediction of small frequencies is usually most susceptible to details of the model.

The tight-binding model reproduces the continuity of the highest frequency, \Fc, deep into the CDW phase and matches the evolution of its effective mass. The large hole pocket associated with \Fc\ does not hybridise immediately upon entering the CDW. Only at  $P\leq\SI{2}{\GPa}$ do the electron pockets hybridise  with the hole band $\gamma$, because only below this pressure the CDW gap overcomes the energy difference and separation in $k$-space between $\gamma$ and $\epsilon$. The hybridisation at $P\approx\SI{2}{\GPa}$ drives the  disappearance of \Fb\ and \Fc, and marks a Lifshitz transition. Thus, our tight-binding model confirms and identifies the experimental Lifshitz transition at $\SI{2}{GPa}$.

The onset of superconductivity at \SI{2}{GPa} is observed to coincide exactly with the emergence of large electron and hole pockets around $\Gamma$. These pockets originate from $L$ and $\Gamma$ respectively, and are connected by the CDW wave vector $\vec{Q}$ in the unreconstructed Brillouin zone. The abrupt emergence of superconductivity at this Lifshitz transition therefore suggests a close relationship between the superconducting and CDW order, as well as a likely interband character for its Cooper pairs. This tangible possibility for the emergence of interband superconductivity in \TiSe\ under pressure raises the question whether the same mechanism may also be at play in Cu-intercalated \TiSe.

\begin{figure*}
\includegraphics[width=.8\textwidth]{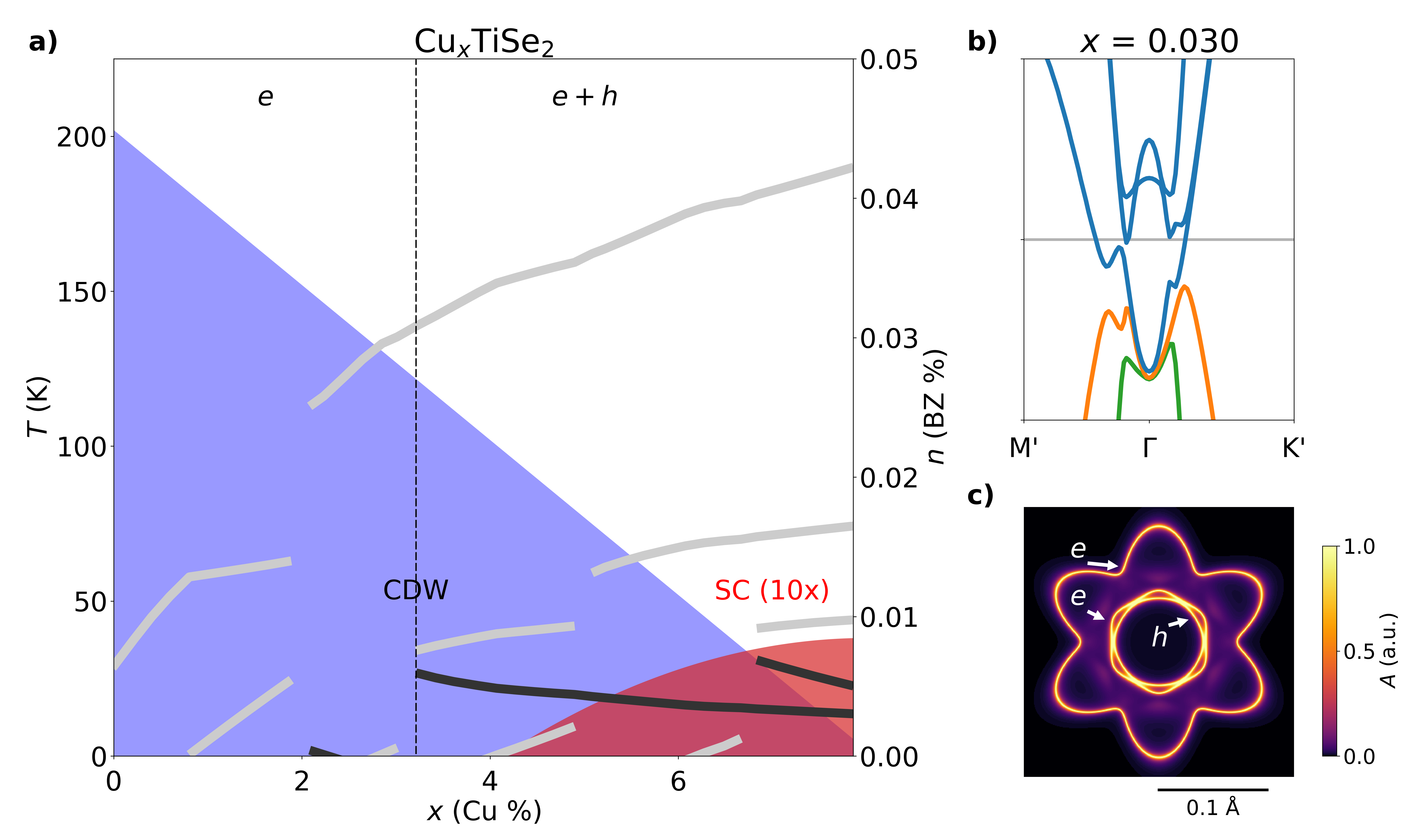}
\caption{\textbf{Doping evolution of the Cu$_x$\TiSe\ Fermi surface}. (a) Calculated evolution of the carrier concentration (right axis) for the electron (grey) and hole (black) Fermi surface pockets, as a function of copper intercalation $x$. Blue and red shaded areas mark the experimental transition temperatures (left axis) of the CDW and superconductivity, respectively , as reported in Ref. \cite{Morosan2006}. (b) Electronic structure and Fermi surface topology illustrating emergence of hole pocket when crossing the Lifshitz transition at $x\approx\SI{3}{\percent}$. Blue indicates the Ti-3$d$ electron band back-folded from the $L$ point whilst green and orange mark the Se-4$p$ hole bands. (c) Total spectral weight at the Fermi level showing the reconstructed Fermi pockets at the same doping as panel (b). The full pockets are visible as a result of finite broadening of the bands for the illustration of spectral weight.
}
\label{fig:doping}
\end{figure*}

\subsection{Doping dependence}

In order to model the Fermi-surface evolution of Cu-intercalated \TiSe, we investigate the doping dependence in our tight-binding model and compare with experimental results by Morosan et al. \cite{Morosan2006}. In our tight-binding model, the ambient-pressure semimetallic electronic structure is established from the match to quantum oscillation measurements (see Supplementary Information E for further discussion of the semimetallic versus semiconducting ground state). The Cu intercalation is  modeled by raising the chemical potential to reflect an electron doping of 0.45 electrons per Cu, as determined from supercell DFT \cite{Jishi2008} and consistent with Hall effect and ARPES measurements \cite{Wu2007, Zhao2007}. The doping leads to the growth of the electron pockets originating from the $L$ point (compare \autoref{fig:doping}(b) with \autoref{fig:theory}(a)). Combined with the suppression of the CDW gap $\Delta(x)$ (scaled to the experimental $\TCDW(x)$ \cite{Morosan2006, Kogar2017a, Kitou2019} as before), this leads to the occurrence of two similar Lifshitz transition as observed under pressure.
 
The first Lifshitz transition is predicted for Cu intercalation at $x\approx\SI{3}{\percent}$, in close proximity to the experimentally observed onset of superconductivity at $x\approx \SI{4}{\percent}$ \cite{Morosan2006}. In \autoref{fig:doping}, we plot the carrier concentrations for individual electron and hole bands (grey and black lines). The electron concentration grows with Cu intercalation as expected. At low $x<\SI{3}{\percent}$ only a single electron pocket is present because the CDW gap is large and the separation of the underlying electron pockets from $L$ and hole pockets from $\Gamma$ is small. Above $\SI{3}{\percent}$ the CDW gap becomes too small to overcome the growing separation and a hole pocket emerges marking a Lifshitz transition (\autoref{fig:doping}). The proximity of the Lifshitz transition with the onset of superconductivity suggests that superconductivity in \CuTiSe\ is dependent on the presence of both electron and hole bands and thus also of interband character.

\section{Discussion}

We have established that superconductivity exists in a dome around the pressure-induced CDW QCP in \TiSe\ with $\Tcmax=\SI{2.9}{K}$. This observation unifies the pressure and doping phase diagram \cite{Morosan2006} and suggests superconductivity is mediated by the CDW fluctuations which soften at the CDW QCP under both tuning parameters. Soft fluctuations near a QCP can manifest as an enhancement and divergence of the effective masses of electrons. In \TiSe, we observe a general enhancement of about a factor 2 over the DFT masses testifying the presence of interactions across the phase diagram. For one frequency, \Fa, we observe the mass enhancement increasing towards the CDW QCP up to a factor 3 over the tight-binding model (\autoref{fig:comparison}), providing evidence for a moderate enhancement of interactions at the CDW QCP. Combined with the evidence for a peak in electron-phonon coupling deduced from the minimum in the exponent of the temperature dependence of the resistivity \cite{Moulding2022}, this indicates the CDW fluctuations play a role in the peak of the superconductivity. These fluctuations may involve phonons, electrons, excitons or a combination. This highlights that the peak of the superconductivity is related to the CDW QCP similar to other quantum critical superconductors.

In addition, for both high-pressure and Cu-intercalated \TiSe, we established the coincidence of a Lifshitz transition with the onset of superconductivity. In both cases we conclude that the presence of electron and hole pockets originating from $L$ and $\Gamma$ respectively are a prerequisite, suggesting that the superconductivity is of interband nature, with pairing enabled by the presence of a CDW with wave vector $\vec{Q}$. A previous renormalisation group study found that interband pairing mechanisms are repulsive in \TiSe\ \cite{Wezel2011}. This implies that the interband superconductivity in \TiSe\ is likely  of unconventional $s_{\pm}$ type in which the superconducting order parameter changes sign between the electron and hole pockets. The sensitivity of superconductivity to strain discussed above is consistent with the sign reversal of $s_{\pm}$ superconductivity. The emergence of superconductivity in \TiSe\ thus closely parallels the established mechanism for interband $s_{\pm}$ superconductivity in the iron pnictides, but is the first system where $s_{\pm}$ pairing is promoted by a CDW.

These results are relevant well beyond the study of \TiSe. 
For instance, the Lifshitz transition in  Ba(Fe$_{1-x}$Co$_x$)$_2$As$_2$ results in  electron and hole pockets connected by the $Q$ vector of the SDW fluctuations at the onset of $s_{\pm}$-superconductivity \cite{Liu2010}, closely resembling the situation discussed here. Likewise, superconductivity in magic-angle trilayer graphene is known to be bounded by Lifshitz transitions \cite{Park2021}. In quasi-1D Bechgaard salts, superconductivity is similarly driven by increasingly mismatched nesting under pressure  \cite{Bourbonnais2011}. In the heavy fermion compound URhGe the magnetic field-induced superconductivity has been proposed to result from a Lifshitz transition \cite{Yelland2011}. 
Across this broad range of physical systems with varying charge and spin interactions considered for Cooper pair formation, Lifshitz transitions demarcating the onset of superconductivity are surprisingly universal. This includes systems with or without divergent effective masses, with a prototypical Fermi liquid dispersion or in the presence of Dirac cones, and in quasi-1D, 2D or 3D.
The specific case of \TiSe\ marks the first material in this class for which charge order, as opposed to spin order, underlies the emergence of a superconducting dome in the phase diagram. Our results suggest that Lifshitz transitions are an overlooked prerequisite for the emergence of unconventional superconductivity at the edge of DW phases in other materials, especially concerning regions of co-existence.

\section*{Methods}
\subsection{High-pressure quantum oscillation measurements}
 
High-pressure measurements used moissanite anvil cells with a culet size of $\SI{800}{\mu m}$. Metallic gaskets were prepared by indenting $\SI{450}{\mu m}$ thick BeCu to approximately $\SI{60}{\mu m}$ followed by drilling a $\SI{450}{\mu m}$ hole. These were insulated with a mixture of alumina and 1266 stycast, which was cured at high pressure between the anvils such that the total thickness of the gasket was less than  $\SI{100}{\mu m}$. A  $\SI{400}{\mu m}$ hole was then drilled in the insulation to act as the sample chamber. Six bi-layer electrodes were deposited on the anvil by firstly argon sputtering  $\SI{20}{\mu m}$ of nichrome followed by thermally evaporating  $\SI{150}{\mu m}$ of gold; excess nichrome was removed with TFN etchant. Finally, gold electrodes were deposited directly onto the sample which were electrically connected to the electrodes on the culet with H21D silver epoxy, which was left to cure at room temperature to avoid degrading the sample.

 The pressure medium used was a 1:1 mixture of pentane and isopentane which is hydrostatic until  $\SI{7.4}{GPa}$ \cite{Klotz2009}. The pressure was determined at room temperature by ruby fluorescence from multiple ruby flakes within the sample chamber. The uncertainty in the pressure was taken as the standard deviation between all the rubies across the sample chamber, both before and after the measurement.

 At the HFML, Nijmegen, a helium-3 cryostat was used to achieve a lowest temperature of  $\SI{0.35}{K}$. In cell 2 where the measurements were performed, the maximum magnetic field applied was $\SI{35}{T}$ and the field was swept at different rates below  $\SI{60}{mT/s}$; the slowest rate used was $\SI{10}{mT/s}$ in order to resolve the highest frequency oscillations. A four-probe method with SR830 lock-in amplifiers was used to measure the resistance, and a Keithley 6221 applied a maximum current of  $\SI{0.5}{mA}$ to avoid heating the sample. Either a model 1900 or SR544 transformer in conjunction with a SR560 pre-amplifier were used to amplify the measured voltage. Effective masses were determined from Lifshitz-Kosevich fits to the resistance between 26 and 35 T for all pressures and frequencies.

\subsection{DFT calculations}

We use WIEN2k with spin-orbit coupling in the Perdew-Burke-Ernzerhof basis set. First, we calculate the pressure dependence of the lattice parameters. This is done by fitting the equation of state to the evolution of the total energy as a function of unit cell volume. The total energy is obtained by relaxing the internal degrees of freedom for fixed unit cell volume. In particular, we minimize the total energy to find the $z$ and $c/a$ parameters for each unit cell volume. Here, $zc$ is the distance of the Se atoms out of the plane. Subsequently, we fit the Birch-Murnaghan equation of state to the evolution of energy versus unit cell volume to convert the volume to pressure. Example fits and a comparison between lattice parameter evolution and X-ray diffraction under pressure are shown in Supplementary Information A. For the doping evolution of the ambient pressure electronic structure, see Supplementary Information B.

In the absence of a CDW above 5 GPa we unambiguously assign the quantum oscillations with the DFT results. We use band shifts up to 150 meV to get the best match between DFT and experiment, see Supplementary Information A for details. We maintain a slight electron doping throughout and although the band shifts reduce the semimetallic overlap compared to natural DFT, we obtain clear semimetallic character at all pressures. We note that quasi-2D neck orbits of the hole bands at the $A$ point predicted by DFT are not detected in quantum oscillations, see Supplementary Information B. 

\subsection{Implementation of the CDW}

To implement the CDW, we fit a tight binding model to the DFT bandstructure as detailed in Supplementary Information C. We restrict ourselves to the $k_z=0$ plane at $\Gamma$ and the $k_z=\pi/2$ plane at $L$ within $\pm$ $\SI{150}{meV}$ of the Fermi level. Subsequently, we implement a CDW reconstruction using the following Hamiltonian:

\begin{equation}
\hat{H}\left(\vec{k}\right)=
    \begin{pmatrix}
        E^{4p}_{\vec{k}} & 0 & \Delta & \Delta & \Delta \\
        0 & E^{4p'}_{\vec{k}} & \Delta & \Delta & \Delta \\
        \Delta & \Delta & E^{3d}_{\vec{k}-{\vec{Q}}} & 0 & 0\\
        \Delta & \Delta & 0 & E^{3d}_{\vec{k}-R\left({\vec{Q}}\right)} & 0\\
        \Delta & \Delta & 0 & 0 & E^{3d}_{\vec{k}-R^2\left({\vec{Q}}\right)})
    \end{pmatrix}.
\end{equation}

Here, $\vec{Q}$ indicates the vector from $\Gamma$ to one of the $L$ points. $R$ denotes rotation under 120 degrees with respect to the $c$-axis to cover the three inequivalent $L$ points. The CDW gap $\Delta$ is approximated as independent of $\vec{k}$ or orbital. This choice was made because no orthogonality exists between out-of-plane Se 2$p_z$ orbitals and in-plane Ti 3$d$ orbitals, unlike the $p_{x,y}$ orbitals relevant for e.g. monolayers of the sister compound TiTe$_2$ \cite{Antonelli2022}, see Supplementary Information C and E. We add a small band-independent linear-in-$P$ energy shift of $\pm$12 meV/GPa to all bands keeping the bands unchanged at \PCDW\ such that the band overlap reduces at low pressure, which was applied to fit the ambient pressure quantum oscillation frequency. 

We now outline our choice for gap size at ambient pressure. BCS theory predicts $\Delta=A k_B\TCDW $ with $A=1.764$ in the weak coupling limit and to our knowledge typically no higher than 3 in the strong-coupling regime. ARPES results commonly estimate a 100 meV separation between conduction and valence bands in the CDW phase, starting from a semiconducting gap of 75-150 meV \cite{Watson2019, Zhao2007, Monney2012a}. We thus use $\Delta(0)=\SI{40}{\milli\electronvolt}$ as a reasonable estimate of the CDW order parameter (observed separation in the CDW minus the pre-existing semiconducting gap), which we find to reproduce the Lifshitz transition at 2 GPa, and to be in accordance with \TiSe\ being in the strong-coupling regime (A $\approx$ 2.3). Under pressure we scale $\Delta(P)=\Delta(0)*\TCDW(P)/\TCDW(0)$ \cite{Moulding2022}. Using these steps, we obtain a model with minimal degrees of freedom that can account for the CDW. 

In summary, after calibrating the DFT to quantum oscillation measurements above \PCDW, we consider only two  parameters  in our tight binding model: the magnitude of the CDW gap $\Delta(0)$ and the linear pressure shift of the bands. These two parameters are determined by two fixed points: the size of the Fermi surface at ambient pressure and the pressure at which the Lifshitz transition is observed. All other quantitative agreement follows naturally, including $\Fc(p)$, the presence of \Fb\ and its pressure dependence, and the persistence of \Fa. These are independent confirmations for our model. The CDW bandstructure shown in \autoref{fig:theory} may be topological as the bands are inverted.


\section*{Additional information}

Data are available at the University of Bristol data repository.

\section*{References}
\bibliography{bibfile}
\end{document}